\documentclass[pra, showpacs, twocolumn, floatfix]{revtex4}

\usepackage{graphicx}
\usepackage{amsmath, amsfonts, amssymb, bm}
\usepackage[]{psfrag}

\psfragscanoff
\def\la{\langle}
\def\ra{\rangle}

\def\ga{\gamma}
\def\gaz{\gamma_2}
\def\gad{\gamma_3}

\def\im{\textrm{i}}

\def\aee{|1\ra \la 1|}
\def\azz{|2\ra \la 2|}
\def\add{|3\ra \la 3|}
\def\aed{|1\ra \la 3|}
\def\ade{|3\ra \la 1|}
\def\wnorm{\: [\ga]}
\def\figref{Fig. }
\def\eqnref{Eq. }
\def\eqnsref{Eqn. }

\begin{document}

\title{Narrow Spectral Feature In Resonance Fluorescence With A Single Monochromatic Laser Field}
\author{J\"org \surname{Evers}}
\email{evers@physik.uni-freiburg.de}
\author{Christoph H. \surname{Keitel}}
\email{keitel@physik.uni-freiburg.de}
\affiliation{Theoretische Quantendynamik, Fakult\"at f\"ur Physik, Universit\"at Freiburg, Hermann-Herder-Stra{\ss}e 3, D-79104 Freiburg, Germany}
\date{\today}

\begin{abstract}
We describe the resonance fluorescence spectrum of an atomic three-level system where two of the states are coupled by a single monochromatic  laser field. The influence of the third energy level, which interacts with the two laser-coupled states only via radiative decays, is studied in detail. For a suitable choice of parameters, this system gives rise to a very narrow structure at the laser frequency in the fluorescence spectrum which is not present in the spectrum of a two-level atom. We find those parameter ranges by a numerical analysis and use the results to derive analytical expressions for the additional narrow peak. We also derive an exact expression for the peak intensity under the assumption that a random telegraph model is applicable to the system. This model and a simple spring model are then used to describe the physical origins of the additional peak. Using these results, we explain the connection between our system, a three-level system in V-configuration where both transitions are laser driven, and a related experiment which was recently reported. 
\end{abstract}


\maketitle

\section{Introduction}
There have been numerous contributions to study the various aspects of the resonance fluorescence of laser-driven few-level atomic systems. One topic of particular interest is the appearance of narrow lines in the fluorescence spectra of laser-driven three-level systems. In these systems, the additional third atomic state together with its coupling to the other two states allows for substantial changes in the resonance fluorescence spectrum, as compared to the spectrum of a two-level atom \cite{mollow}. In \cite{schmal1} the appearance of sub-natural line widths was predicted for a three-level system in V-configuration, a prediction which was verified experimentally \cite{gauthier}. Also in three-level systems in $\Lambda$ - configuration the possibility of line narrowing was found \cite{lambda}. In \cite{zhu95}, the spontaneous decay between two atomic levels was modified by an additional atomic level coherently coupled to one of the other states. One of the results was a possible occurrence of narrow lines in the fluorescence spectrum. \cite{zhou97} found narrow lines due to quantum interference in an atomic system with two close energy levels which couple to the same vacuum modes. Line narrowing can also be found in systems with more than three energy levels. \cite{paspalakis} reports spectral narrowing controllable by the phase difference between two driving laser fields in a four level system. In principle it is also possible to have unbounded line narrowing in the emission spectrum with essentially complete transfer of the corresponding intensity into the narrow feature \cite{keitel99}. 
Somewhat similar, also the probe absorption of an atomic few-level system may be altered favourably under the influence of driving laser fields (See e.g. \cite{absorption}).

A different mechanism for the appearance of narrow lines is electron shelving \cite{shelving, shelving-exp}. In \cite{hp95, garraway}, this phenomenon was discussed in a three-level atomic system driven by two laser fields in a V - configuration. One of the fields is assumed to be much weaker than the other. This allows for a shelving of the atomic population into the weakly coupled atomic level. Because of the long lifetimes of such shelved states, the fluorescence spectrum may exhibit narrow peaks. Compared to the usual two-level fluorescence spectrum, the spectrum of the atomic system discussed in \cite{hp95, garraway} contains an additional very narrow peak centered at the laser frequency for a suitable choice of parameters. The peak is controllable to some extent by altering the strength of the weak coupling laser.

\begin{figure}
\psfrag{O}[][l]{$\bm \Omega$}
\psfrag{g}{$\bm \ga$}
\psfrag{g2}{$\bm \gaz$}
\psfrag{g3}{$\bm \gad$}
\psfrag{1}{$\bm |1\rangle$}
\psfrag{2}{$\bm |2\rangle$}
\psfrag{3}{$\bm |3\rangle$}
\includegraphics[height=3.5cm]{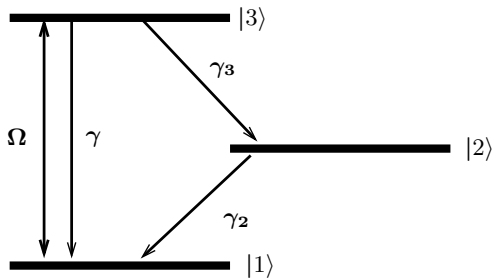}
\caption{\label{pic-system} The system of interest in bare state representation. $\Omega$ is the Rabi frequency of the driving laser field, and $2\ga$, $2\gaz$ and $2\gad$ are the total decay widths.}
\end{figure}

Recently there has been an experiment testing the ansatz of electron shelving \cite{buehner00}. In the experiment a three-level atomic system different from the one described above was used, in which only one transition is driven by a laser field. The third atomic level was coupled to the driven levels by radiative decays as shown in \figref \ref{pic-system}. Experimentally, this was realized by a setup as in \figref \ref{pic-4level}. In the experiment, the additional narrow peak due to an electron shelving was found. But although the atomic system used was different, the experimental results were reported to be consistent with the theoretical results in \cite{hp95}.

In this paper, we provide the theoretical background for the atomic system used in the experiment mentioned above. We set up the equations of motion and calculate the resonance fluorescence spectrum. By a numerical analysis, we find suitable parameter ranges for the exhibition of the additional narrow peak, which may be useful for further experiments. We use these results to derive analytical expressions for the additional peak in an approximation which is similar to the secular limit for high laser intensities. Also we derive an exact analytical expression for the peak intensity valid for any choice of parameters under the assumption that a random telegraph model is applicable to describe the physical origin of the narrow peak. Then we use this random telegraph model to clarify the connection between the experiment and the two different atomic systems in a discussion of the physical origins of the additional narrow peak. Finally, we present a simple spring model to give an intuitive picture of the atomic dynamics. This model is able to reproduce the different components found in the resonance fluorescence spectrum and allows to easily derive approximate expressions for their widths.

The paper is structured as follows. In the second section we derive expressions for the resonance fluorescence spectrum of the given atomic system. In section three, we use the derived equations for a numerical analysis of the resonance fluorescence spectra. The results of this analysis motivate the approximations made in the first part of the following analytical treatment of the problem. In the second part of the analytical section, we use the telegraph model to derive expressions for more general parameter ranges. In section four we give two different physical interpretations for the occurrence of the additional narrow feature in the fluorescence spectra, the random telegraph model and a spring model. The first model is used to clarify the connection between our system of interest, the system where both transitions are driven coherently and the experiment. The second model is the spring model for an intuitive picture of the atomic dynamics. Section five will discuss and summarize the results.

\section{Preliminary considerations}
In this section, the atomic system studied in this paper is described. We set up the equations of motion and use these to derive expressions for the resonance fluorescence spectrum.

\subsection{System of interest}
\begin {figure}[t]
\psfrag{O}[][l]{$\bm \Omega$}
\psfrag{g}{$\bm \ga$}
\psfrag{g2}{$\bm \gaz$}
\psfrag{g3}{$\bm \gad$}
\psfrag{1}{$\bm |1\rangle$}
\psfrag{2}{$\bm |2\rangle$}
\psfrag{3}{$\bm |3\rangle$}
\psfrag{4}{$\bm |4\rangle$}
\psfrag{R}{\textbf{R}}
\includegraphics[height=3.5cm]{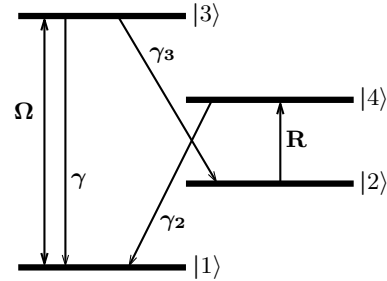}
\caption{\label{pic-4level} Possible experimental realization of the atomic system of interest. R represents a pumping from $|2\rangle$ to $|4\rangle$.}
\end{figure}

The atomic system considered in this paper is shown in \figref \ref{pic-system}.
$\hbar \omega_i \; (i=1,2,3)$ are the respective energies of the three levels, and $2\gamma$, $2\gamma_2$ and $2\gamma_3$ are the decay constants (Einstein coefficients). The transition $1 \leftrightarrow 3$ with the frequency $\omega _0 = (\omega _3 - \omega _ 1)$ is driven by a laser with the frequency $\omega _L$ and the Rabi frequency $\Omega$. $\Delta=(\omega _L - \omega _0)$ is the detuning. In the analytical calculations, we take 
\begin{equation}
\gamma _2, \gamma _3 \ll \gamma \label{gen-assumption}
\end{equation}
as a general assumption as this will turn out to be the most interesting parameter range. We neglect all terms of order higher than $\gamma _i/\gamma\: (i=2,3)$ in the analytical calculations if not noted otherwise. Thus the second atomic level is only weakly coupled to the other levels. In an experiment this is either possible by using a quadrupole transition for $3 \leftrightarrow 2$ or $1 \leftrightarrow 2$ or by using dipole transitions including a fourth ancilla level as shown in \figref \ref{pic-4level}. In this figure, R is a pumping mechanism which transfers atomic population from $|2\rangle$ to $|4\rangle$.


\subsection{Dynamical equations}
The Hamiltonian $H$ for the system is
\begin{eqnarray*}
H &=& H_F + H_I \\
H_F &=& \hbar (\omega _1 -\Delta) \aee + \hbar \omega _2 \azz  + \hbar \omega _3 \add \\
H_I &=& \hbar \Delta \aee - \frac{\hbar \Omega}{2}(\aed e^{\im \omega_L t} + \text{ h.c.}) 
\end{eqnarray*}
The Hamiltonian $V$ in the interaction picture with respect to $H_F$ is
\begin{equation}
V = \hbar \Delta \aee - \frac{\hbar \Omega}{2} \left (\ade + \aed \right ) \label{eqn-hamilton}
\end{equation}
The equations of motion for the corresponding density matrix $\rho$ in the interaction picture are
\begin{eqnarray}
\dot{\rho} _{33} &=& -2(\ga + \gad)\rho_{33} + \frac{\im \Omega}{2}(\rho _{31} - \rho_{13}) \label{eom-anfang} \\
\dot{\rho} _{11} &=& 2\ga \rho_{33} + 2\gaz \rho_{22} - \frac{\im \Omega}{2}(\rho _{31} - \rho_{13})\\
\dot{\rho} _{31} &=& \frac{\im \Omega}{2}(\rho_{33}-\rho_{11}) - (\ga + \gad - \im \Delta)\rho_{31} \label{eom-ende}
\end{eqnarray}
with $\rho _{22} = 1- \rho _{33} - \rho _{11}$ and $\rho _{ij} = (\rho _{ji})^*$ for $i,j\in \{1,2,3\}$. $\rho _{32}$ and $\rho _{21}$ relax with the rate $(\ga + \gad + \gaz)$ and $\gaz$ respectively. Introducing $a=\gamma_2/\gamma_3$, the steady state solution $\rho ^{ss}$ to this system becomes 
\begin{eqnarray}
\rho ^{ss}_{33} &=& \frac{1}{N} a\Omega^2 \nonumber\\
\rho ^{ss}_{11} &=& \frac{1}{N} a (4\Delta^2 + 4\ga^2 + 8\ga \gad + 4\gad ^2 + \Omega^2) \nonumber \\
\rho ^{ss}_{22} &=& \frac{1}{N} \Omega^2 \nonumber \\
\rho ^{ss}_{31} &=& -\frac{1}{N} 2a\Omega (\Delta + \im (\ga+\gad))  \nonumber \\
\rho ^{ss}_{32} &=& \rho ^{ss}_{21} = 0  \label{steady-state}
\end{eqnarray}
with the abbreviation
\[
N = \Omega^2 + 2a(2\Delta^2+2\ga^2+4\ga \gad + 2\gad ^2 + \Omega^2) 
\]
Upon inserting $a=\gaz / \gad$ and setting $\gad$ to zero in the above expressions, the result simplifies to the usual expressions for a laser-driven two-level system. Throughout the paper, a superscript ``2-level'' denotes values for this two-level system, whereas a superscript of ``3-level'' or no superscript means the corresponding values for the three-level system discussed here.

\subsection{Observables of interest}
The equations of motion \eqnsref (\ref{eom-anfang})-(\ref{eom-ende}) can be written as
\begin{eqnarray*}
\frac{d}{dt}\vec{\rho}  &=& \bm B \: \vec{\rho} + \vec{I} \\
\vec{\rho} &=& (\rho_{33}, \rho_{11}, \rho_{31}, \rho_{13})^T \\
&=& \langle \vec{\sigma} \rangle \\
\vec{\sigma} &=& (\sigma _1, \sigma _2, \sigma _3, \sigma _4)^T \\
  &=& (\add, \aee, \ade, \aed)^T 
\end{eqnarray*}
with time independent time evolution matrix ${\bm B}$ and a constant vector $\vec{I}$ given by 
\begin{eqnarray*} 
\bm{B}=\left ( 
\begin {array}{cccc}
-2\: (\ga+\gad) & 0 & \frac{\im}{2}\Omega & - \frac{\im}{2}\Omega \\
2\: (\ga-\gaz) & -2\: \gaz & -\frac{\im}{2}\Omega &  \frac{\im}{2}\Omega \\
\frac{\im}{2}\Omega & - \frac{\im}{2}\Omega & -\im \Delta - \ga - \gad & 0 \\
- \frac{\im}{2}\Omega & + \frac{\im}{2}\Omega & 0 & \im \Delta - \ga - \gad \\
\end {array}
\right ) 
\end{eqnarray*}
and $\vec{I} = (0, 2\gaz, 0, 0)^T$.

The normalized resonance fluorescence spectrum $S(\omega)$ is then given by the real part of the Fourier transformation of the two-time correlation function $\langle \sigma_3(t)\sigma_4(t')\rangle$ as \cite{ct75}
\begin{equation}
S(\omega) = \frac{1}{\pi \rho^{ss}_{33}} \text{Re}\: \int ^{\infty}_0 e^{-\im \omega (t-t')} \langle \sigma_3(t)\sigma_4(t')\rangle\:
dtdt' \label{calc-full-spec}
\end{equation}
The normalization is chosen such that the integral of $S(\omega)$ over all frequencies $\omega$ yields unity. Thus the normalized expressions for the different contributions to the total resonance fluorescence spectrum are equal to the relative intensities of the spectral components.

Following \cite{ct75}, we split the expression for $\sigma$ into its average value and the deviation from the average
\begin{equation*}
\sigma_i = \langle \sigma _i \rangle + \delta \sigma _i
\end{equation*}
for $i\in \{1,2,3,4\}$. By inserting this, the correlation function can be separated into two terms corresponding to its average motion and its fluctuations around the average:
\begin{equation}
\langle \sigma_3(t)\sigma_4(t')\rangle = \langle \sigma_3(t)\rangle \langle \sigma_4(t')\rangle + \langle \delta \sigma_3(t) \delta \sigma_4(t')\rangle \label{calc-spec-split}
\end{equation}
The first term corresponding to the average motion yields the coherent or elastic part $S_{coh}(\omega)$ of the normalized spectrum 
\begin{equation}
S_{coh}(\omega) = \frac{|\rho^{ss}_{31}|^2}{\rho^{ss}_{33}} \: \delta(\omega - \omega_L) \label{full-el-spec}
\end{equation}
Thus we have
\begin{equation}
I_{\text{coh}} = \frac{|\rho^{ss}_{31}|^2}{\rho^{ss}_{33}} \label{elast-int}
\end{equation}
as the relative intensity of the elastic peak.

The second term represents the incoherent or inelastic part of the spectrum. According to the quantum regression theorem we have for $t'>t$ and $i\in \{1,2,3,4\}$
\begin{equation}
\frac{d}{dt}\langle \vec{\sigma}(t')\sigma_i(t) \rangle = {\bm B} \: \langle \vec{\sigma}(t')\sigma_i(t) \rangle + \vec{I} \: \langle \sigma_i(t) \rangle
\end{equation}
This together with \eqnsref (\ref{calc-full-spec}) and (\ref{calc-spec-split}) yields for the total normalized incoherent spectrum $S_{inc}(\omega)$ consisting of the Mollow spectrum and the additional narrow peak
\begin{equation}
S_{inc}(\omega) = \frac{1}{\pi \rho^{ss}_{33}}\: \text{Re}\: \left ( \frac{1}{\im \omega - {\bm B}}\vec{R} \right )_4 \label{full-inel-spec} 
\end{equation}
where the brackets with the subindex mean that the fourth component has to be taken and ``Re'' indicates that the real part of the following expression has to be taken. Also we have 
\begin{eqnarray*}
\vec{R} &=& \langle \delta \vec{\sigma}(0) \delta \sigma_{4}(0)\rangle^{ss} \\
&=& \langle \vec{\sigma}(0)\sigma_{4}(0)\rangle^{ss} - \langle \vec{\sigma}(0) \rangle^{ss} \langle \sigma_{4}(0) \rangle^{ss} 
\end{eqnarray*}
where the index ``$ss$'' denotes the steady state value.

\section{Calculation of the spectra}
In this section we calculate the resonance fluorescence spectra 
emitted by the given atomic system. First the equations derived in the previous chapter are used for a numerical analysis without approximations.
In the second part, we derive analytical expressions for one of the spectral components using the results of the numerical analysis. 

\subsection{\label{sec-numerical} Numerical analysis}

\begin {figure}[t]
\psfrag{xlbl}[tl]{$\omega-\omega_L \wnorm$}
\psfrag{ylbl}[tc][c][1][180]{relative intensity}
\includegraphics[width=7cm]{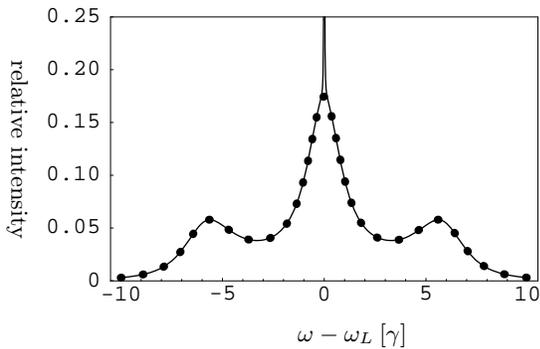}
\caption{\label{pic-full-spk} Sample spectrum. Parameters used are $\ga = 1$, $\Omega = 6$, $\gad=0.005$, $a = 0.3$, $\Delta = 0$. The solid line shows the full normalized inelastic spectrum. The dotted line shows the normalized Mollow spectrum extracted from the full spectrum as shown in section \ref{sec-secular}.}
\end{figure}
\begin {figure}[t]
\psfrag{xlbl}[tl]{$\omega-\omega_L \wnorm$}
\psfrag{ylbl}[ct][c][1][180]{relative intensity}
\includegraphics[width=7cm]{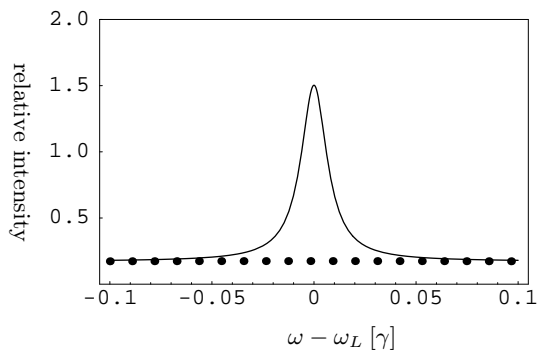}
\caption{\label{pic-peak} Closeup on the additional narrow peak. Plots and parameters are as in \figref \ref{pic-full-spk}. Note the very different axes scales.}
\end{figure}

The sum of the elastic part in \eqnref (\ref{full-el-spec}) and the inelastic part in \eqnref (\ref{full-inel-spec}) give the full normalized resonance fluorescence spectrum without approximations other than those adopted in the derivation of the equations of motion. In this section we use this expression for a numerical analysis of the spectra emitted by the system of interest. In all plots, the elastic component is omitted. Thus narrow peaks centered at the laser frequency are always new structures due to the addition of the third atomic level.

A sample spectrum is shown in \figref \ref{pic-full-spk}. The solid line shows the full normalized inelastic spectrum. The dotted line shows the normalized Mollow spectrum extracted from the full spectrum. As elaborated later, this Mollow spectrum is the spectrum one would expect from a driven two-level system \cite{mollow}. The additional narrow peak at $\omega _L$ can easily be seen. A closeup on this peak with very different scales on both axes is shown in \figref \ref{pic-peak}. The additional peak can be distinguished from the coherent peak due to its finite width.

Since the elastic and the additional narrow peak are both centered at the laser frequency, experimental observation of the narrow peak is a challenging matter. In recent experiments, optical heterodyne detection was used to measure fluorescence spectra with sufficiently high resolution\cite{buehner00, heterodyne}.

\begin {figure}[t]
\psfrag{lblxxxxx}[lt][B][1][-43]{$\omega - \omega_L  \wnorm$}
\psfrag{lblzzzzz}[t][b][1][-83]{\begin{minipage}[t]{2.5cm}\center rel. intensity [arb. units]\end{minipage}}
\psfrag{lbla}[][B][1][0]{(a)}
\psfrag{lblb}[][B][1][0]{(b)}
\psfrag{lblc}[][B][1][0]{(c)}
\psfrag{lbld}[][B][1][0]{(d)}
\includegraphics[width=7cm]{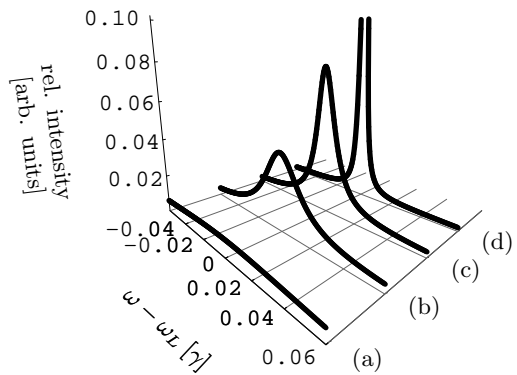}
\caption{\label{pic-num-g3} Normalized inelastic spectrum for different values of $\gad$. $a=\gaz/\gad$ is kept fixed at $a=0.3$. The other parameters are $\ga = 1$, $\Omega =  8$, $\Delta = 0$. (a) $\gad = 0.05$; (b) $\gad = 0.01$; (c) $\gad = 0.005$; (d) $\gad = 0.001$. The z-axis has arbitrary units and a pedestal due to the Mollow peak was removed.}
\end{figure}
\begin {figure}[t]
\psfrag{lblxxxxx}[tl][B][1][-43]{$\omega - \omega_L  \wnorm$}
\psfrag{lblzzzzz}[t][b][1][-83]{\begin{minipage}[t]{2.5cm}\center rel. intensity [arb. units]\end{minipage}}
\psfrag{lbla}[][B][1][0]{(a)}
\psfrag{lblb}[][B][1][0]{(b)}
\psfrag{lblc}[][B][1][0]{(c)}
\psfrag{lbld}[][B][1][0]{(d)}
\includegraphics[width=7cm]{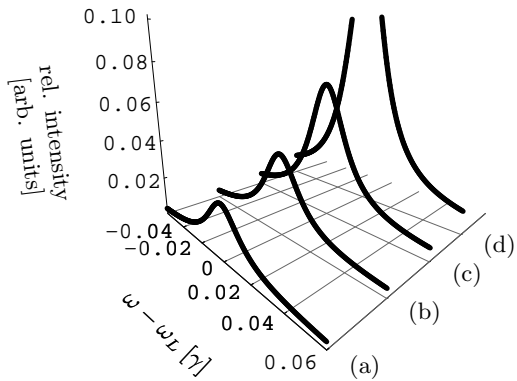}
\caption{\label{pic-num-w} Normalized inelastic spectrum for different values of $\Omega$. The other parameters are $\ga = 1$, $\gad= 0.01$, $a = 0.3$, $\Delta = 0$. (a) $\Omega = 10$; (b) $\Omega = 8$; (c) $\Omega = 6$; (d) $\Omega = 4$. The z-axis has arbitrary units and a pedestal due to the Mollow peak was removed.}
\end{figure}

\figref \ref{pic-num-g3} shows spectra obtained for different values of $\gad$. $\gaz$ is adjusted according to the fixed ratio $a = \gaz / \gad$, and also the other parameters are kept fixed. Effectively, the figure shows the influence of the coupling strength of the additional atomic level to the laser-driven ones on the additional peak. Only a very narrow frequency range around the laser frequency is plotted. The $z$-axis has arbitrary units, and the contribution from the Mollow peak centered at the laser frequency has been subtracted. It can be seen that for large $\gad$, i.e. strong coupling of the additional atomic state to the driven ones, the additional narrow peak becomes broader and seems to vanish. For decreasing $\gad$, the peak amplitude increases while its width decreases. As long as \eqnref (\ref{gen-assumption}) is satisfied, the relative peak intensity is unaffected by the value of $\gad$ if $a$ and the other parameters are kept fixed. Therefore in experiments it is possible to choose $\gad$ and $\gaz$ as small as possible compared to $\ga$ in order to have a narrow peak with large amplitude. This justifies our assumption \eqnref (\ref{gen-assumption}) for the analytical part later.

A comparison of spectra for different Rabi frequencies $\Omega$ is shown in \figref \ref{pic-num-w}. Again, only a narrow frequency range around the laser frequency is shown, and the z-axis has arbitrary units with a removed Mollow contribution as in the previous plot. The narrow peak at $\omega _L$ vanishes for increasing $\Omega$, just as the elastic component not shown in the picture does. This shows that both the elastic and the additional narrow component are non-secular features. If the Rabi-frequency is chosen too large, their contribution to the complete spectrum will be small.

\begin {figure}[t]
\psfrag{xlbl}[][][1][-20]{$\Omega \wnorm$}
\psfrag{ylbl}{$\Delta \wnorm$}
\psfrag{zlbl}[t][Br][1][-84]{\begin{minipage}[t]{2.5cm}\center abs. intensity [arb. units]\end{minipage}}
\includegraphics[width=6.6cm]{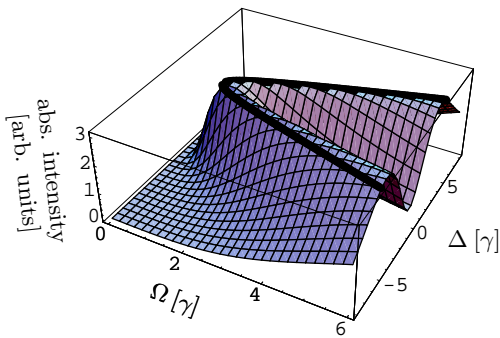}
\caption{\label{pic-elast-int} Absolute intensity of the narrow peak component of the spectrum against $\Omega$ and $\Delta$. Parameters used in the plot are $\ga = 1$,  $\gad = 0.005$, $a = 0.3$. The solid lines show the detuning calculated using the condition in \eqnref  (\ref{det-saddle}).}
\end{figure}

\begin {figure}[t]
\psfrag{xlbl}[tl]{$\omega-\omega_L \wnorm$}
\psfrag{ylbl}[ct][c][1][180]{relative intensity}
\includegraphics[width=7cm]{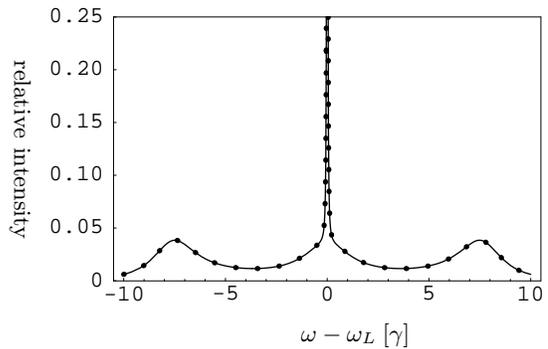}
\caption{\label{pic-sec-spk} Sample spectrum with $\Delta = \Delta _ {\text{max}}$. The other parameters used are as in \figref \ref{pic-full-spk}. The solid line shows the full normalized inelastic spectrum. The dotted line shows the sum of the extracted normalized Mollow spectrum and the approximation for the narrow peak in \eqnref (\ref{approx-peak}).}
\end{figure}

\begin {figure}[t]
\psfrag{xlbl}[tl]{$\omega-\omega_L \wnorm$}
\psfrag{ylbl}[ct][c][1][180]{relative intensity}
\includegraphics[width=7cm]{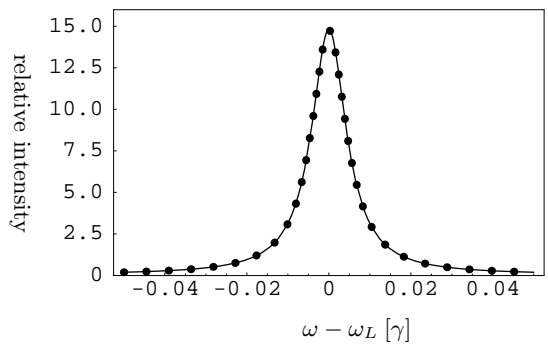}
\caption{\label{pic-sec-peak} Closeup on \figref \ref{pic-sec-spk}. Note the very different axes scales.}
\end{figure}

It is however possible to choose an appropriate detuning $\Delta$ to have large intensity in the narrow peak even for very large Rabi frequencies. This can be seen from \figref \ref{pic-elast-int} which shows the absolute intensity of the narrow peak plotted against the Rabi frequency and the detuning. For any fixed value of $\Delta$, the peak intensity vanishes for increasing Rabi frequency $\Omega$. However by appropriately changing $\Delta$ with $\Omega$ it is possible to move along the saddle of maximum intensity in \figref \ref{pic-elast-int}. In \figref \ref{pic-sec-spk} and \figref \ref{pic-sec-peak}, compared to \figref \ref{pic-full-spk} and \figref \ref{pic-peak} only the detuning has been changed to a suitable value $\Delta_{\text{max}}$ estimated from \figref \ref{pic-elast-int}. Nevertheless the peak amplitude has increased by more than one order of magnitude, while the width at half height has only decreased by a factor of approximately two thirds. Thus the peak intensity with the appropriately chosen detuning is more than 7 times larger than in the case with zero detuning. The exact values for the amplitudes $A$, the widths $\Gamma$ and the intensities $I$ of the narrow peak in the plots are:
\begin{eqnarray*}
\frac{A(\Delta = \Delta_{\text{max}})}{A(\Delta=0)} &\approx& \frac{14.8}{1.3} \: \approx \: 11.1 \\
\frac{\Gamma(\Delta = \Delta_{\text{max}})}{\Gamma(\Delta=0)} &\approx& \frac{0.0051}{0.0077}  \: \approx  \: 0.7  \\
\frac{I(\Delta = \Delta_{\text{max}})}{I(\Delta=0)} &\approx& \frac{0.076}{0.010}  \: \approx  \: 7.4
\end{eqnarray*}

This is a clear indication for the fact that for zero detuning, the Rabi frequency chosen for the plots is already large enough to account for a secular suppression of the peak intensity.

Later we will derive an analytical expression for a suitable detuning $\Delta _{\text{max}}$. In the limit $\Omega \gg \gamma, \: \ga \gg \gamma_3, \gamma_2$ and $\Delta=\Delta _{\text{max}}$ (see \eqnref (\ref{sec-approx})), modifying $\Omega$, $\gamma$ or $\gamma_3$ and $\gamma_2$ with constant $a$ by one order of magnitude typically changes the absolute intensity of the peak by less than one percent. The absolute intensity of the additional peak only depends on $a=\gaz / \gad$.  It reaches its maximum value for $a=(2\sqrt{3})^{-1}$. For this value of $a$, the relative intensity of the additional peak is approximately $24\%$. Therefore this limit is suitable for further experiments and might even be used to measure the relation between $\gad$ and $\gaz$.

\begin {figure}[t]        
\psfrag{lblxx}[t][B][1][0]{$\omega - \omega_L \wnorm$}
\psfrag{lblyy}[t][c][1][180]{relative intensity}
\psfrag{lbla}[][c][1][0]{(a)}
\psfrag{lblb}[][c][1][0]{(b)}
\psfrag{lblc}[][c][1][0]{(c)}
\psfrag{lbld}[][c][1][0]{(d)}
\includegraphics[width=7cm]{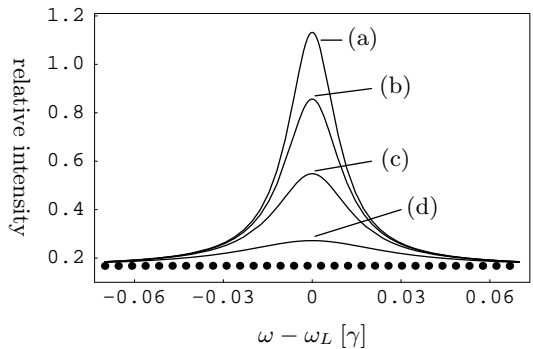}
\caption{\label{pic-num-a} Normalized inelastic spectrum for different values of $a$. The other parameters are $\ga = 1$, $\gad= 0.01$, $\Omega = 8$, $\Delta = 0$. (a) $a = 0.01$; (b) $a = 0.1$; (c) $a = 0.3$; (d) $a = 1$. The dotted line shows the corresponding two-level Mollow spectrum.}
\end{figure}

The dependence of the narrow peak on $a$ is shown in \figref \ref{pic-num-a}. In all the curves, $\gad$ is kept fixed, and $\gaz$ is adjusted according to $\gaz = a \gad$. Thus this figure shows the dependence of the narrow peak on the relation of the couplings of the additional atomic level to the upper and to the lower laser-driven level. The dotted line shows the Mollow spectrum of the corresponding two-level system as a reference. For increasing $a$, the peak width increases and the peak height decreases. In contrast to the independence on $\gad$, the peak intensity does depend on $a$. Therefore it is not possible to decrease the peak width by choosing very low values for $a$, as this would be equivalent to very low values for $\gaz$. A low value for $\gaz$ however traps the population in the additional atomic state and thus yields a low total fluorescence intensity. The value $a=0.3$ as chosen in most of our plots is close to the value which yields a maximum absolute peak intensity of the narrow peak.

\begin {figure}[t]        
\psfrag{lblxx}[t][B][1][0]{$\omega - \omega_L \wnorm$}
\psfrag{lblyy}[t][c][1][180]{relative intensity}
\includegraphics[width=7cm]{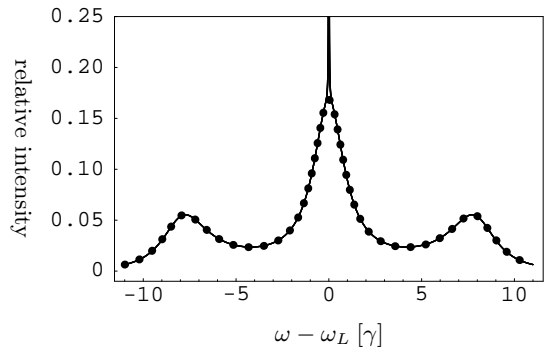}
\caption{\label{pic-num-a-ganz} Normalized inelastic spectrum for different values of $a$. The parameters are chosen as in \figref \ref{pic-num-a}. The dotted line shows the two-level Mollow spectrum. The four solid lines show spectra for the system discussed here. Apart from the narrow peak, the different curves are almost on top of each other.}
\end{figure}

The Mollow contributions to the inelastic spectrum are almost unaffected by changing $a$ as shown in \figref \ref{pic-num-a-ganz}. While \figref \ref{pic-num-a} only shows a very narrow frequency region around the laser frequency, this figure depicts the full inelastic spectrum for the same sets of parameters as in \figref \ref{pic-num-a}. The dotted spectrum for the two-level system and the four three-level spectra only differ in the small frequency region shown in \figref \ref{pic-num-a}, where the additional narrow peak dominates. A similar plot shows that the Mollow spectrum also does not depend on the value of $\gad$ and $\gaz$ as long as \eqnref  (\ref{gen-assumption}) is fulfilled.

\subsection{\label{sec-analytical} Analytical considerations}
In this subsection, we derive analytical expressions for the spectral contributions by the additional narrow peak using two different methods.
The results of the numerical analysis encourage us to focus on a restricted domain of parameters. Thus in the first method, we assume \eqnref  (\ref{gen-assumption}) and choose the Rabi frequency to be large as compared to the spontaneous emission widths (secular limit). In this parameter range, approximations to simplify the results are possible. In the second method, we calculate an exact expression for the intensity of the peak under the assumption that a random telegraph model can be applied to describe the origin of the narrow peak. This assumption will be justified in section \ref{sec-telegraph}.

\subsubsection{\label{sec-secular} Secular approach}
In order to obtain an analytical expression for the additional peak, a Lorentzian shape is assumed, because the usual Mollow peaks are also approximately Lorentzian in the parameter ranges where they are well separated. In the Mollow case, this separation is equivalent to a high intensity of the driving laser field. This assumption allows for the parametrization of the peak only by the height and the width at half height. To calculate the peak height, we expand \eqnref (\ref{full-inel-spec}) with respect to $\gamma_3 / \gamma \ll 1$. The zeroth order corresponds to the usual Mollow spectrum without the additional peak as shown in \figref \ref{pic-full-spk}. The first order corresponds to the additional peak. Thus the peak height $A$ is given by the difference of the full spectrum and the Mollow spectrum at $\omega = \omega _L$.
We find
\begin{equation}
A=\frac{2(\Delta^2 + \gamma^2)\Omega^2}{(\Omega^2 + 2a(2\Delta^2 + 2\gamma^2 + \Omega^2))^2 \: \pi \: \gamma _3} \label{secular-ampl}
\end{equation}
as an approximation for the peak amplitude $A$ which is valid if the condition in \eqnref (\ref{gen-assumption}) is satisfied.

The peak width can be calculated without further approximation from the real part of the corresponding eigenvalues of the time evolution matrix $\bm B$ of the system 
\cite{ct75}, but the resulting expression is too complicated to be shown here. It turns out to be difficult to simplify the expressions for the width of the additional peak with parameters chosen at the maximum peak 
intensity, where $\Omega$, $\gamma$ and $\Delta$ are all approximately the same. It is however possible to make the Rabi frequency $\Omega$ very large without loosing much of the peak intensity, if one appropriately changes the detuning $\Delta$ with the laser intensity. This can easily be done in experiments. This approach is similar to the secular limit usually made in explaining the Mollow spectrum in the dressed state picture. However the elastic component and the additional peak are not secular features and vanish for increasing Rabi frequencies if the other parameters are kept fixed.   The reason for this is that the driven atomic transition saturates such that the atomic dipole the laser field couples to vanishes. An adequate choice of $\Delta$ effectively prevents this saturation. We choose $\Delta$ such that the peak amplitude on the right hand side (rhs.) of \eqnref (\ref{secular-ampl}) has a maximum value:
\begin{equation}
\Delta ^2_{\text{max}} = \frac{(1+2a)\Omega^2 - 8a\gamma^2}{8a} \label{det-saddle}
\end{equation}
This is the analytical expression for the detuning $\Delta _{\text{max}}$ for which we already obtained a numerical value by reading off from \figref \ref{pic-elast-int} in section \ref{sec-numerical}. Our choice is justified by the fact that the variation of the amplitude in the parameter range around the maximum intensity values are larger than the variations of the width.
The absolute peak intensity is plotted against the Rabi-frequency and the detuning in \figref \ref{pic-elast-int}. For any fixed value of $\Delta$, the intensity vanishes for large Rabi frequencies $\Omega$. The condition \eqnref (\ref{det-saddle}) is plotted as the solid lines on top of the saddle in the picture. It determines the detuning $\Delta$ such that for any given value of $\Omega$, the absolute peak intensity is close to its maximum value. 
In the approximation
\begin{equation}
\Omega \gg \gamma,\: \Delta = \Delta _{\text{max}}, \: \ga \gg \gamma_3, \gamma_2  \label{sec-approx}
\end{equation}
as explained above we find
\begin{eqnarray}
\Gamma_{\text{peak}}^{\text{sec}} &=& \frac{6a(1+2a)}{1+6a} \gamma_3  \label{breite-approx}\\
A_{\text{peak}}^{\text{sec}} &=&  \frac{1}{9a(1+2a)\: \pi \: \gamma_3}\\
I_{\text{peak}}^{\text{sec}} &=&  \pi \cdot \Gamma_{\text{peak}}^{\text{sec}} \cdot A_{\text{peak}}^{\text{sec}}\\
&=& \frac{2}{3(1+6a)}\label{int-approx}
\end{eqnarray}
for the width $\Gamma_{\text{peak}}^{\text{sec}}$, the amplitude $A_{\text{peak}}^{\text{sec}}$ and the intensity $I_{\text{peak}}^{\text{sec}}$ of the narrow peak. Thus in this limit, the intensity of the additional peak only depends on $a$. A change of $\gamma _2$ and $\gamma_3$ with constant $a$ only alters the shape of the peak, whereas the other parameters do not have any effect. These results agree with and explain the results obtained by the numerical analysis.

With the parameters in \eqnsref (\ref{breite-approx}) - (\ref{int-approx}), the contribution $S_{peak}(\omega)$ of the narrow peak to the normalized fluorescence spectrum is
\begin{equation}
S_{peak}(\omega) = \frac{I_{\text{peak}}^{\text{sec}}}{\pi}  \frac{\Gamma_{\text{peak}}^{\text{sec}}}{(\omega - \omega _L)^2 + (\Gamma_{\text{peak}}^{\text{sec}})^2 }\label{approx-peak}
\end{equation}
Both the exact spectrum and a spectrum obtained using the above approximations are shown in \figref \ref{pic-sec-spk}. A closeup on the frequency range around $\omega = \omega_L$ is plotted in \figref \ref{pic-sec-peak}. In both pictures, the solid line is given by the exact expression for the normalized inelastic spectrum in \eqnref (\ref{full-inel-spec}). The dotted lines were calculated by adding the rhs. of \eqnref (\ref{approx-peak}) to the two-level Mollow spectrum obtained as explained above. It can be seen that the approximate parameters in \eqnsref  (\ref{breite-approx}) - (\ref{int-approx}) are in very good agreement with the exact results.

\subsubsection{\label{sec-physical} Random telegraph approach}
This subsection differs from the other parts of this paper in that it does not try to argue that the random telegraph model is a good description of the physics of the additional narrow peak. Instead, it uses the model as the starting point for further calculations assuming that it is applicable. This will be justified in section \ref{sec-telegraph}. Nevertheless, the fact that the results we will derive agree with other results in the respective limits gives further support for the model.

Starting from the random telegraph model, the intensity of the additional peak can be calculated in a different way. According to the model, the additional peak stems from the stochastic modulation of the elastic component in the spectrum due to long dark and light periods in the resonance fluorescence signal. These modulations partially broaden the elastic component, but do not  affect the relative weights of the inelastic Mollow components. The modulations only shift relative intensity from the elastic peak to the additional narrow peak. Therefore the sum of the relative intensity of the elastic component and of the additional peak has to be the same as the relative intensity of the elastic component in a system without these modulations. Thus the relative intensity of the additional peak $I_{\text{peak}}^{\text{phys}}$ is equal to the difference of the relative intensities of the elastic component in a two-level system $I_{\text{elastic}}^{\text{2-level}}$ and the corresponding value $I_{\text{elastic}}^{\text{3-level}}$ of the system discussed here:
\begin{equation}
I_{\text{peak}}^{\text{phys}} = I_{\text{elastic}}^{\text{2-level}} - I_{\text{elastic}}^{\text{3-level}}
\end{equation}
The intensities of the elastic components are easy to calculate as in \eqnref (\ref{elast-int}) using the steady state values in \eqnref  (\ref{steady-state}) of the system and the corresponding values for the two-level system.
Under the assumption that the random telegraph model is applicable, this yields as an exact expression for the relative intensity of the additional peak
\begin{equation}
I_{\text{peak}}^{\text{phys}}=\frac{2(\Delta^2+\gamma^2-4a\gamma\gamma_3 - 2a\gamma _3^2)\Omega^2}{K(\Omega^2 +2a(K  + 4\gamma\gamma_3 + 2\gamma_3 ^2 ))} \label{phys-int} 
\end{equation}
where $ K= (2\Delta^2 + 2\gamma^2 + \Omega^2)$. This result is valid for any possible choice of parameters. 

If \eqnref (\ref{gen-assumption}) is satisfied, a series expansion with respect to $\gad/\ga$ up to zeroth order yields as an approximation for the peak intensity $I_{\text{peak}}^{\text{phys}}$:
\begin{equation}
I_{\text{peak}}^{\text{phys}} \approx \frac{2(\Delta^2+\ga^2)\Omega^2}{K(\Omega^2 + 2aK)}\label{simp-phys-int}
\end{equation}
Then the expression in \eqnref (\ref{secular-ampl}) for the peak amplitude can be used to calculate an approximate expression for the peak width $\Gamma_{\text{peak}}^{\text{phys}}$:
\begin{equation}
\Gamma_{\text{peak}}^{\text{phys}} = \frac{\Omega^2 + 2a(2\Delta^2 + 2\gamma^2 + \Omega^2)}{2\Delta^2 + 2\gamma^2 + \Omega^2} \: \gad \label{phys-width}
\end{equation}
Thus \eqnsref (\ref{secular-ampl}), (\ref{simp-phys-int}) and (\ref{phys-width}) describe the amplitude, the intensity and the width of the peak if \eqnref (\ref{gen-assumption}) is fulfilled. In the limit in \eqnref (\ref{sec-approx}), these results simplify to the previous results \eqnsref (\ref{breite-approx}) - (\ref{int-approx}). 

This method of calculating an exact expression for the peak intensity is not restricted to the atomic system discussed here. It can be used in any system where the stochastic modulations due to light and dark states account for a partial broadening of the elastic peak. For example, it is easy to derive an exact expression for the peak intensity in the atomic system driven by two lasers only by subtracting two steady state values using this technique. In the respective limit, this expression simplifies to the approximate result given in \cite{hp95}, Eq. (41).

\section{Physical origin of the additional peak}
It is of course of great interest why there is an additional structure in the resonance fluorescence spectrum compared to a two-level system. In several previous publications, a random telegraph model was used to explain the dynamics in atomic systems like the one discussed here \cite{shelving,shelving-exp,hp95,garraway, jump,porrati89,javanaien86,telegraph-dressed,pegg,buehner00, hp96, rmp70}. This model is applied to the present atomic system in the first subsection \ref{sec-telegraph}. In the second subsection \ref{sec-spring}, a spring model is used to give an intuitive picture of the dynamics of the atomic system.
 
\subsection{\label{sec-telegraph} Random telegraph model}
If the fluorescence light of a system exhibiting the additional narrow peak is recorded by a broadband photo-detector, the signal shows a random sequence of bright and dark periods. Bright periods consist of a closely spaced sequence of fluorescence pulses, whereas dark phases are long periods in which no photons are emitted. The bright phases correspond to the usual atomic dynamics between the two levels connected by the driving laser field as it is also present in a two-level system. The dark phases correspond to a trapping of the atomic population in a metastable state which in general is a coherent superposition of the atomic states. Because of the abrupt changes between light and dark phases, the fluorescence light behaves like a random telegraph signal.

In \cite{hp95}, this random telegraph model was used to explain the occurrence of an analogous additional peak in a different atomic system. In their system, both the transitions $1\leftrightarrow 3$ and $1\leftrightarrow 2$ are coherently driven by lasers, but no transitions from level 3 to 2 are allowed ($\gad=0$). The laser driving the transition $1\leftrightarrow 2$ is assumed to be weak compared to the other laser, and $\gaz$ is set to zero during the calculations. Thus in the system in \cite{hp95}, the additional atomic energy level $|2\rangle$ is coherently coupled to one of the other levels, whereas in our system it is incoherently coupled to both driven levels by radiative decays. The result in \cite{hp95} was that the random fluctuations of the fluorescence light due to the change between light and dark states partially broaden the coherent spectral component. This broadening is essentially determined by the average length of the bright and dark periods.

The fact that only a part of the elastic component becomes broadened can easier be understood if one thinks of an ensemble rather than of a single atom. In fact, a different approach to calculate the spectra has to be taken to describe single atom spectra \cite{hp96, springer70}. In an ensemble, at any given time there is always a certain proportion of atoms in a light period. Because all the coherently emitted light is in phase with the incident laser, the sum of all these contributions forms the elastic part of the total fluorescence. This can be seen using the analytical expressions for the average dark period length $\tau_D$ and the average bright period length $\tau_B$ which will be derived in section \ref{sec-duration}. If \eqnref (\ref{gen-assumption}) is satisfied, we have from \eqnsref (\ref{elast-int}), (\ref{simp-phys-int}), (\ref{tau-d}) and (\ref{tau-b}) 
\begin{eqnarray*}
\alpha &:=& \frac{\tau_B}{\tau_D+\tau_B} = \frac{I_{\text{coh}}}{I_{\text{peak}}^{\text{phys}} + I_{\text{coh}}} \\
\beta &:=& \frac{\tau_D}{\tau_D+\tau_B} = \frac{I_{\text{peak}}^{\text{phys}}}{I_{\text{peak}}^{\text{phys}} + I_{\text{coh}}} 
\end{eqnarray*}
where ($I_{\text{peak}}^{\text{phys}} + I_{\text{coh}}$) is the relative intensity of the coherent peak in an unbroadened  system not exhibiting the additional narrow peak. This intensity is equal to the relative elastic peak intensity in a two-level system. Thus a single atom is the proportion $\alpha$ of the time in a light period. Because of the phase coupling of all coherently emitted light to the driving laser field, in a large ensemble, this proportion of the total number of atoms can be considered as radiating continuously. Therefore effectively the proportion $\alpha$ of the unbroadened elastic intensity remains in the elastic part of a partially broadened spectrum. The remaining part $\beta$ of the ensemble effectively can be seen as never having a bright period, but only dark periods. Thus the proportion $\beta$ of the unbroadened elastic intensity is shifted to the additional peak. 

There has also been an experiment measuring the fluorescence spectrum of single Yb(171) and Yb(172) ions \cite{buehner00}. The spectrum of the first isotope exhibits dark and light periods, whereas the second has a continuous spectrum. The atomic system used in this experiment is not the system presented in \cite{hp95}, but the system described here with a fourth ancilla level (see \figref \ref{pic-4level}). Still, the experimental results were reported to be consistent with the theory in \cite{hp95}.  

The mechanism for the exhibition of light and dark periods is not the same in the two atomic systems. In the system driven by two lasers, population coherently moves between the fluorescent transition and the additional atomic state. Thus a dark state is not characterized by all the population being trapped in the additional atomic level, but by a stable superposition of all three atomic levels. This can be seen by the fact that a dark phase can end by an emission from both the upper and the metastable state \cite{porrati89}. In the system presented here, all the population is trapped in the additional state during a dark period; dark periods can only end by a transition from this level \cite{javanaien86}. Because there are two different mechanisms which lead to analogous observations, it is reasonable that it is only the common fact that there are light and dark periods which accounts for the additional structure in the spectrum. Therefore the random telegraph model should be a good theoretical model whenever a system exhibits long bright and dark periods, no matter what the specific atomic structure is  \cite{rmp70}.

As it seems, the experimental results were compared to results derived from the random telegraph model. The results of this model only depend on the average length of dark and light periods, $\tau _D$ and $\tau _B$. These parameters were not calculated using the experimental parameters, but measured during the experiment. However when $\tau _D$ and $\tau _B$ are measured, one does not take into account the specific structure of the given atomic system, but only tests the random telegraph model. On this level of description, both the atomic systems with coherent and with incoherent coupling of the third atomic level are equivalent. This explains the results in \cite{buehner00}. As the measured spectra are in quantitative agreement with the predictions from the telegraph model, the experiment justifies the usage of this model.

One way of testing the theoretical description of a specific atomic system is to calculate the spectrum without using the telegraph model and to compare it to the measured spectra. Another way is to calculate $\tau _D$ and $\tau _B$ using a quantum jump approach as in \cite{hp95}. Then the calculated durations can be compared to the measured ones. This will be done in paragraph \ref{sec-duration}.

There is another argument for the random telegraph model. If this model is correct, then the fluctuations in the spectrum should not change the relative intensity of the Mollow spectrum, but only shift intensity from the elastic peak to the new narrow peak. A comparison of the relative intensity of the coherent part of a two-level system and of the sum of coherent and peak intensity in the system discussed here shows that this is true up to small errors due the approximations made to derive the intensity of the additional peak. Using this physical interpretation for the origin of the additional peak, one can calculate the peak intensity by taking the difference of the two-level coherent intensity and the three-level coherent intensity, which can both be calculated without approximations. (see section \ref{sec-physical})

The random telegraph model also allows for a dressed state description of the additional peak. In \cite{telegraph-dressed} the occurrence of long dark and light periods is explained in a dressed state picture. As the additional narrow peak stems from these long average durations, this description can be seen as a valid explanation for the narrow peak.

\subsubsection{\label{sec-duration} Calculation of $\tau _D$ and $\tau _B$}
To calculate the average durations of bright and dark periods, we follow a method proposed in \cite{pegg}. At the time $t=0$, we assume the atom not to be in the additional atomic level, i.e. $\rho_{22}(0) = 0$. We need to find the expectation value $\tau_B$ for the time the atom stays in the driven subspace $\{|1\rangle, |3\rangle \}$ without moving into $|2\rangle$. If at a later time $t_1$ the atom is still in the driven subspace, the life expectancy at that time is again $\tau_B$. If $P_{31}(t_1)$ is the probability of remaining in the driven subspace without moving into $|2\rangle$ during the time $t_1$, we have
\begin{equation}
\tau_B = P_{31}(t_1)(t_1+\tau_B) + \left (1-P_{31}(t_1) \right)ft_1
\end{equation}
where $0\leq f \leq 1$. The first term represents the case in which the atom remains in the driven subspace. The second term is the probability of not staying in that subspace multiplied by the time $ft_1$ at which the atom moves out of the driven subspace. Thus we have
\begin{equation}
\tau_B = \frac{t_1}{1-P_{31}(t_1)}-t_1(1-f)
\end{equation}
For large $t_1$, the atom may jump out of the subspace and back into it. This stops the current bright period, but does not change $\rho_{22}$. For $t_1\rightarrow 0$ however the probability of leaving the subspace is dominated by single jumps into the additional atomic level. Therefore we have 
\begin{equation}
\lim _{t_1\rightarrow 0}  (1-P_ {31}(t_1)) = \lim _{t_1\rightarrow 0} \rho_{22}(t_1) 
\end{equation}
and we 
obtain
\begin{equation}
\tau_B = \lim_{t_1\rightarrow 0} \frac{t_1}{\rho_{22}(t_1)}
\end{equation}
and thus
\begin{equation}
\tau_B^{-1}=\left ( \frac{d\rho_{22}}{dt}\right )_{t=0} \label{duration-b}
\end{equation}
with the condition $\rho_{22}(0)=0$. A similar argumentation for the time the atom continuously stays in $|2\rangle$ yields 
\begin{equation}
\tau_D^{-1}=- \left ( \frac{d\rho_{22}}{dt}\right )_{t=0} \label{duration-d}
\end{equation}
with $\rho_{22}(t=0)=1$.

From the equations of motion \eqnsref (\ref{eom-anfang})-(\ref{eom-ende}) we have for $\rho_{22}$:
\begin{equation*}
\dot{\rho}_{22} = 2\gad \rho_{33} - 2\gaz \rho_{22}
\end{equation*}
This yields for the average length of a dark period using \eqnref (\ref{duration-d})
\begin{eqnarray}
\tau_D^{-1} &=& - \biglb ( 2\gad \rho_{33} (0) - 2\gaz \rho_{22} (0) \bigrb ) \nonumber \\
&=& 2\gaz \label{tau-d}
\end{eqnarray}
where we have used $\rho_{22}(t=0)=1$ and thus $\rho_{33}(t=0)=0$. This result is what one would expect, as the system is in the additional atomic level during a dark phase. Thus the length of such a dark period is given by the inverse of the total decay width $2\gaz$ of this level.

For the average length of a bright period we obtain from \eqnref (\ref{duration-b}) under the condition $\rho_{22}(t=0)=0$:
\begin{eqnarray*}
\tau_B^{-1} &=&  2\gad \rho_{33} (0) - 2\gaz \rho_{22} (0) \\
&=& 2\gad \rho_{33} (0)
\end{eqnarray*}
As the atom only evolves between the driven atomic levels, and as we look at the atomic system long after the laser was switched on when it has reached its quasistationary state, we can replace $\rho_{33} (0)$ by the steady state value for $\rho_{33}$ in the case of a two-level system. Using $a=\gaz / \gad$ and setting $\gad$ to zero in \eqnref (\ref{steady-state}) yields:
\begin{equation*}
\rho_{33}^{ss, 2-\text{level}} = \frac {\Omega^2}{2(2\Delta^2+2\ga^2+\Omega^2)}
\end{equation*}
Thus the result is
\begin{equation}
\tau_B^{-1} = \frac {\Omega^2}{2(2\Delta^2+2\ga^2+\Omega^2)} \: 2\gad \label{tau-b}
\end {equation}
As one would expect, the average length of a bright period is inversely proportional to the total decay width $2\gad$ out of the laser driven subspace.

Equations \eqnsref (\ref{tau-d}) and (\ref{tau-b}) show that our general assumption in \eqnref (\ref{gen-assumption}) of a weak coupling of the additional atomic level to the driven levels is equivalent to long average dark and bright phases.


\begin{figure}[t]
\psfrag{g}{$\gamma$}
\psfrag{g3}{$\gamma _3$}
\psfrag{g2}{$\gamma _2$}
\psfrag{w}{$\hbar \widetilde{\Omega}$}
\psfrag{d}{$\hbar \Delta$}
\psfrag{l}{$\hbar \omega _L$}
\psfrag{3n}[rB]{$|3, N\rangle$}
\psfrag{1n}[rB]{$|1, N\rangle$}
\psfrag{2n}{$|2, N\rangle$}
\psfrag{2nm}{$|2, N-1\rangle$}
\psfrag{3nm}[rB]{$|3, N-1\rangle$}
\psfrag{1np}[rB]{$|1, N+1\rangle$}
\psfrag{pn}{$|+, N+1\rangle$}
\psfrag{mn}{$|-, N+1\rangle$}
\psfrag{pnm}{$|+, N\rangle$}
\psfrag{mnm}{$|-, N\rangle$}
\includegraphics[width=6cm]{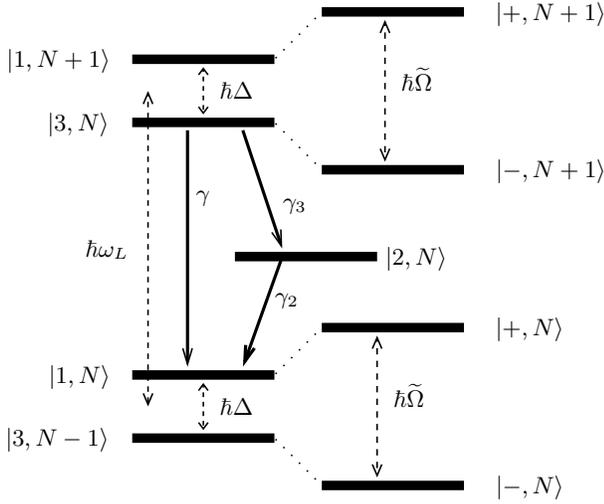}
\caption{\label{pic-dressed-system} Relation between the bare and dressed states in the system of interest. The figure is plotted for the case $\Delta > 0$. The solid arrows show the bare state transition widths; the dashed arrows denote energy splittings. $(N)$, $(N-1)$ and $(N+1)$ are the respective photon numbers in the laser field mode. $\widetilde{\Omega}=\sqrt{\Omega^2+\Delta^2}$ is the generalized Rabi frequency.}
\end{figure}

As shown in \cite{hp95}, one can derive an expression for the width $\Gamma_{\text{telegraph}}^{\text{peak}}$ of the additional narrow peak in terms of the average length of dark and bright periods if these durations are long:
\begin{equation*}
\Gamma_{\text{telegraph}}^{\text{peak}}= \tau_B^{-1}  + \tau_D^{-1}
\end{equation*}
Inserting the rhs. of \eqnsref (\ref{tau-d}) and (\ref{tau-b}) we get
\begin{equation}
\Gamma_{\text{telegraph}}^{\text{peak}} = \frac{\Omega^2 + 2a(2\Delta^2 + 2\gamma^2 + \Omega^2)}{2\Delta^2 + 2\gamma^2 + \Omega^2} \: \gad \label{peak-breite-dauer}
\end{equation}
which is the result we already obtained in \eqnref (\ref{phys-width}). This is not surprising as \eqnref (\ref{phys-width}) was derived under the assumption \eqnref (\ref{gen-assumption}) which guarantees long average lengths of the dark and bright periods.

If the parameters describing the atomic system and the laser field are known, expressions \eqnsref (\ref{tau-d}) and (\ref{tau-b}) can be compared to average light and dark period times obtained by experiments as \cite{buehner00}. Alternatively, \eqnref (\ref{peak-breite-dauer}) may be compared with an experimentally determined width of the narrow feature. This allows to test the quantum optical description of the atomic system via the telegraph model. Equally assuming that this description is appropriate, these results can for example be used to measure the decay widths $\gaz$, $\gad$ or the effective decay width from $|2\rangle$ to $|1\rangle$ of a system of ancilla pumping laser and decay from $|4\rangle$ to $|1\rangle$ as in \figref \ref{pic-4level}.

\subsection{\label{sec-spring} Spring model}
In this section, we will use the picture of a simple spring model to describe the atomic dynamics. To physically interpret the calculated spectrum, we transfer the system to the dressed state picture (c.f. \figref \ref{pic-dressed-system}). The dressed states are defined as the eigenstates of the system Hamiltonian in the interaction picture in \eqnref (\ref{eqn-hamilton}):
\begin{eqnarray*}
|+,\: N\rangle &=& \sin (\theta)\:  |3,\: N-1\rangle +\cos (\theta) \: |1,\: N\rangle \\
|-,\: N\rangle &=& \cos (\theta)\:  |3,\: N-1\rangle -\sin (\theta) \: |1,\: N\rangle 
\end{eqnarray*}
with 
\begin{eqnarray}
\sin (\theta) &=& \frac{\Omega}{\sqrt{2}(\Delta^2+\Omega^2-\Delta \sqrt{\Delta^2 + \Omega^2})^{\frac{1}{2}}} \label{sinval}\\
\cos (\theta) &=& \sqrt{1-\sin^2 (\theta)} \label{cosval}
\end{eqnarray}
$N$ and $N-1$ are the respective photon numbers in the laser field mode. The third bare state $|2\rangle$ is not coupled to the driving laser field and thus remains the same in the dressed state picture.

By choosing a detuning according to \eqnref (\ref{det-saddle}), the Rabi frequency of the driving laser field can be made very large as compared to all transition widths in the system without loosing weight in the elastic part of the spectrum or the additional peak (see \figref \ref{pic-elast-int}). This secular limit allows for a separation of the central components whose time evolution is at the laser frequency $\omega _L$ and of the sideband components.

The resulting equations for the corresponding population vector  $\vec{\varrho}$ of the central frequency parts at $\omega _L$ in secular approximation can be written as
\begin{eqnarray}
\frac{d}{dt} \vec{\varrho} &=& {\bm M} \cdot \vec{\varrho} \label{sec-middle} \\
\vec{\varrho} &=& (\varrho_{++},\varrho_{--}, \varrho_{22})^T \nonumber 
\end{eqnarray}
\begin{figure}[t]
\psfrag{D1}{$D_1$}
\psfrag{D2}{$D_2$}
\psfrag{D3}{$D_3$}
\psfrag{1}{$m_+$}
\psfrag{2}{$m_2$}
\psfrag{3}{$m_-$}
\includegraphics[width=4.8cm]{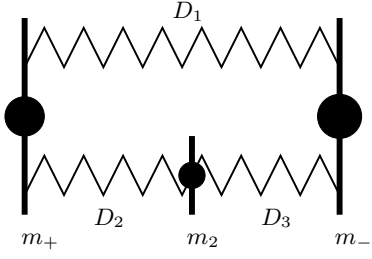}
\caption{\label{pic-spring} The spring system used as a model for the atomic dynamics.}
\end{figure}
$\varrho_{ii}$ for $i\in \{+,-,2\}$ are the populations of the dressed states summed over all possible laser field photon numbers.
\begin{figure}[h]
\psfrag{+}{$+$}
\psfrag{2}{$2$}
\psfrag{-}{$-$}
\psfrag{(a)}{$(a)$}
\psfrag{(b)}{$(b)$}
\psfrag{(c)}{$(c)$}
\includegraphics[width=7cm]{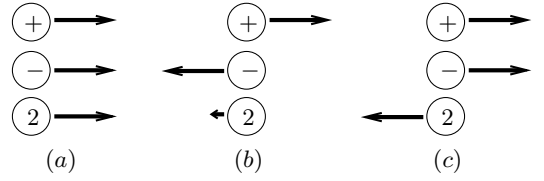}
\caption{\label{pic-modes} Fundamental modes of the spring model. (a) corresponds to the elastic component, (b) to the Mollow peak, and (c) to the additional narrow peak. The arrows show the relative elongation directions of the masses. The small arrow in mode (b) is a correction term only used in the closer analysis.}
\end{figure}

The time evolution matrix $\bm M$ is given by
\begin{widetext}
\begin{eqnarray}
{\bm M} = 
\left ( 
\begin {array}{ccc}
-2\gamma_3 \sin ^2 \theta -2\gamma \sin^4 \theta & 2\gamma \cos^4 \theta & 2\gamma_2 \cos ^2 \theta \\
2\gamma \sin^4 \theta & -2\gamma_3 \cos ^2 \theta -2\gamma \cos^4 \theta & 2\gamma_2\sin ^2 \theta \\
2\gamma_3 \sin^2 \theta & 2\gamma_3 \cos^2 \theta & -2\gamma_2 \\
\end {array}
\right ) \nonumber
\end{eqnarray}
\end{widetext}

An essentially equivalent picture for these spectral components can be obtained using a mechanical spring model as shown in \figref \ref{pic-spring}. The equations of motion for this spring model are
\begin{eqnarray}
\frac{d^2}{dt^2} \vec{Q} &=& {\bm F} \cdot \vec{Q} \label{sec-feder} \\
\vec{Q} &=& (m_+q_+,m_-q_-,m_2q_2)^T 
\end{eqnarray}
with
\begin {eqnarray*}
{\bm F} &=& 
\left ( 
\begin {array}{ccc}
\frac{D_1+D_2}{m_+}& -\frac{D_1}{m_-} & -\frac{D_2}{m_2} \\
-\frac{D_1}{m_+} & \frac{D_1+D_3}{m_-} & -\frac {D_3}{m_2} \\
-\frac{D_2}{m_+} & -\frac{D_3}{m_-} & \frac{D_2+D_3}{m_2} \\
\end {array}
\right )
\end {eqnarray*}
$m_i$ are the masses and $q_i \: \left (i\in\{+,-,2\} \right )$ are the elongations of the masses. $D_i \: (i\in \{1,2,3\})$ are the spring constants as defined in \figref \ref{pic-spring}.

After transforming \eqnref (\ref{sec-middle}) to a second order differential equation in $t$ and identifying the atomic population with the product of mass and elongation, a comparison with the spring model equation \eqnref (\ref{sec-feder}) yields the relations
\begin{eqnarray*}
m_+ &=& \sin ^{-4} (\theta) \\
m_- &=& \cos ^{-4} (\theta) \\
m_2 &=& \frac{\gamma _3}{\gamma _2}\cdot \sin ^{-2} (\theta) \cos ^{-2} (\theta)
\end{eqnarray*}
\begin{eqnarray*}
D_1 &=& 3\gamma ^2 + 4\gamma \gamma _3 -4\gamma _2 \gamma _3 + \gamma ^2 \cos (4\theta) \\
D_2 &=& \frac{\gamma _3}{\sin ^2 (\theta)} \{ \gamma + 4\gamma _2 + 2\gamma _3 \\
&& - 2(\gamma + \gamma _3)\cos (2\theta) + \gamma \cos (4\theta)  \}\\
D_3 &=& \frac{\gamma _3}{\cos ^2 (\theta)} \{ \gamma + 4\gamma _2 + 2\gamma _3 \\
&&+ 2(\gamma + \gamma _3)\cos (2\theta) + \gamma \cos
(4\theta) \}
\end{eqnarray*}
for the masses and the spring constants.

From the above equation for $m_2$ we can see that the parameter $a=\gaz / \gad$ determines $m_2$ which is the mass corresponding to the additional atomic level. This explains why $a$ is a crucial parameter for the system. For example in the secular limit, the peak properties depend on $a$ although all other parameters have only very little influence on the results as shown in the analytical calculations.

\begin {figure}[t]
\psfrag{L+}{$|+\rangle$}
\psfrag{L-}{$|-\rangle$}
\psfrag{L2}{$|2\rangle$}
\psfrag{Z1}[][Br][1][-90]{$2\ga \sin^4(\theta)$}
\psfrag{Z2}[][Br][1][-90]{$2\ga \cos^4(\theta)$}
\psfrag{Z3}[][Br][1][-52]{$2\gad \sin^2(\theta)$}
\psfrag{Z4}[][Br][1][-52]{$2\gaz \cos^2(\theta)$}
\psfrag{Z5}[B][Br][1][52]{$2\gad \cos^2(\theta)$}
\psfrag{Z6}[][Br][1][52]{$2\gaz \sin^2(\theta)$}
\includegraphics[width=5cm]{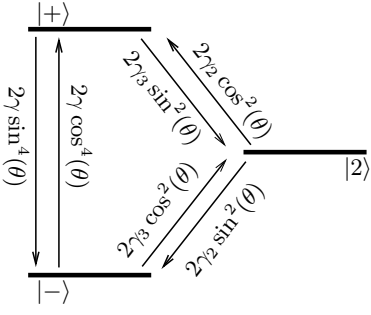}
\caption{\label{pic-uebergang} Possible transitions between the dressed states in the spring model with corresponding rates.}
\end{figure}

There are three fundamental oscillation modes in the spring model, which can be calculated from the eigenvectors of the time evolution matrix ${\bm F}$. See \figref \ref{pic-modes} for a symbolic depiction of the modes. The arrows represent the relative elongation directions of the masses. \figref (\ref{pic-uebergang}) shows the possible transitions between the dressed states with the respective transition widths. We denote the transition rate from $i$ to $j$ as $\Gamma_{ij}$ $\left (i,j\in\{+,-,2\} \right )$.  

The first mode (a) corresponds to a uniform motion of all three masses. As there is no relative movement between the masses, none of the spring constants - or in the atomic model none of the decay constants - are relevant. Thus the total decay width $\Gamma_{\text{a}} ^{\text{spring}}$ in this mode is zero:
\[
\Gamma_{\text{a}} ^{\text{spring}}=0
\]
This mode corresponds to the elastic peak which is of zero width. 

In the second mode (b), the masses $+$ and $-$ oscillate against each other, whereas mass $2$ is approximately at rest. We begin by neglecting the small movement of mass $2$ shown in \figref \ref{pic-modes}, which will be included later in a closer analysis. In this approximation, only the spring constants or decays between $+$ and $-$ are important for the time evolution of the system, which are of  order of $\ga$ (c.f. \figref \ref{pic-uebergang}). Therefore this mode corresponds to the inelastic Mollow peak at $\omega = \omega _L$. The expected decay width $\Gamma_{\text{b}} ^{\text{spring}}$ is the sum of the decay widths from $+$ to $-$ and vice versa:
\begin{eqnarray*}
\Gamma_{\text{b}} ^{\text{spring}} &=& \Gamma _{+-} + \Gamma _{-+}  \\
&=& 2 \ga \sin^4 (\theta) + 2 \ga \cos^4 (\theta) \\
&=& \frac{3+\cos(4\theta)}{2}\: \ga
\end{eqnarray*}
The exact width $\Gamma_{\text{b}}$ of this peak calculated in secular approximation from the equations of motion for our three-level system \eqnref (\ref{sec-middle}) up to first order in $\gad / \ga$ is
\begin{equation}
\Gamma_{\text{b}} =  \frac{3+\cos(4\theta)}{2}\: \ga + \frac{5+3\cos(4\theta)}{3+ \cos(4\theta)}\: \gad \label{sec-exakt-mollow}
\end{equation}
Thus the predicted and the exact widths are the same up to a correction term of first order in $\gad / \ga$.

In the third mode (c), $+$ and $-$ oscillate against $2$. Only the transitions from $+$ and $-$ to $2$ and vice versa are used, but not the ones between $+$ and $-$. As the transitions between $2$ and $+$ or $-$ are of order $\gad$, this mode corresponds to the additional peak. The expected width $\Gamma_{\text{c}} ^{\text{spring}}$ is 
\begin{eqnarray*}
\Gamma_{\text{c}} ^{\text{spring}}&=& \Gamma _{+2} + \Gamma _{-2} + \Gamma _{2+} + \Gamma _{2-}\\
&=&2\gad \sin^2 (\theta) + 2\gad \cos^2 (\theta)\\
&&  + 2\gaz \cos^2 (\theta) +2\gaz \sin^2 (\theta)\\
&=& 2(a + 1)\gad
\end{eqnarray*} 
The exact width $\Gamma _{\text{c}}$ of this peak in secular approximation from \eqnref (\ref{sec-middle}) up to first order in $\gad / \ga$ is
\begin{equation}
\Gamma_{\text{c}} =  2(a + 1)\gad - \frac{5+3\cos(4\theta)}{3+ \cos(4\theta)}\: \gad \label{sec-exakt-peak}
\end{equation}
Also for this mode the widths are the same up to a correction term of first order in $\gad / \ga$.

Using this simple picture, it is possible to predict the widths of the spectral components up to a correction term of order $\gamma_3$ for the last two modes. This is a minor correction for the Mollow peak, which is of order $\gamma$, but can be a large correction to the width of the additional peak, which is also of  order $\gamma_3$. It is interesting to note that the correction term in both modes is the same up to a change in sign. 

A closer analysis shows that the populations of the dressed states in the modes corresponding to the Mollow and the additional peak are not the same. Therefore the relevant transition widths do not contribute equally to the total width of the respective modes. This is the reason for the correction terms found above. Thus in the following, we use the populations of the dressed states in the different fundamental modes as weights for the transition widths and take the small correction due to the movement of mass $2$ in the Mollow peak mode into account. By this it is possible to obtain the exact secular approximation results $\Gamma_{\text{b}}$ and $\Gamma_{\text{c}}$ in first order in $\gad / \ga$ for the widths of the Mollow and the additional peak from the spring model.

For mode (b) corresponding to the central inelastic Mollow  peak we have as relative weights in the dressed states up to first order in $\gad / \ga$ from the corresponding eigenvector:
\begin{eqnarray*}
w_{\text{b}}^{+} &=& \frac{\alpha _{\text{b}}}{2} \\
w_{\text{b}}^{-} &=& \frac{\alpha _{\text{b}}}{2} \left (1 - \frac{m_- - m_+}{m_+ + m_-} \cdot \frac{\gad}{\ga} \right )\\
w_{\text{b}}^{2} &=& \frac{\alpha _{\text{b}}}{2} \cdot \frac{m_- - m_+}{m_+ + m_-} \cdot \frac{\gad}{\ga}
\end{eqnarray*}
The subindex ``b'' denotes the mode, and the superscripts ``+'', ``-'' and ``2'' mark the corresponding masses or dressed states. $\alpha_{\text{b}} = w_{\text{b}}^{+} + w_{\text{b}}^{-} + w_{\text{b}}^{2}$ is a normalization constant. This constant is due to the fact that the eigenvector from which the weights are calculated does not have a definite norm, such that only the ratios between the different weights can be calculated from it. The normalization constant does not have to be unity, as only the total population has to be normalized to unity. This total population is the solution to \eqnref (\ref{sec-middle}) and consists of a linear combination of the contributions from the three different modes, thus allowing for a different normalization of the weights of a single mode.

Clearly the weights should not be larger than unity, because the transition widths cannot contribute with a prefactor larger than one to the total width of the peak. Thus we have $0 \leq \alpha_{\text{b}} \leq 2$. As in a peak due to a single decay the corresponding single transition width is equal to the total width and as additional transitions can only increase the total width, it is plausible that the largest weight is unity. Therefore we assume $\alpha_{\text{b}} = 2$. 

The relative weights of the dressed states in mode (c) corresponding to the additional peak up to zeroth order in $\gad / \ga$ are
\begin{eqnarray*}
w_{\text{c}}^{+} &=& \frac{\alpha _{\text{c}}}{2} \cdot \frac{m_+}{m_+ + m_-}\\
w_{\text{c}}^{-} &=& \frac{\alpha _{\text{c}}}{2} \cdot \frac{m_-}{m_+ + m_-}\\
w_{\text{c}}^{2} &=& \frac{\alpha _{\text{c}}}{2} 
\end{eqnarray*}
Again, the subindex ``c'' denotes the mode. A similar argument as for mode (b) yields $\alpha_{\text{c}} = 2$ as assumption for the normalization constant. It is not necessary to take higher orders in $\gad / \ga$ into account for the narrow peak mode weights as the corresponding transition widths for the mode (c) are already of first order in $\gad / \ga$. 

With these expressions for the weights, we obtain as width $\Gamma ^{w}_{\text{b}}$ of the Mollow peak up to first order in $\gad / \ga$:
\begin{eqnarray*}
\Gamma ^{w}_{\text{b}} &=& w_{\text{b}}^{+} \:(\Gamma _{+-} + \Gamma _{+2}) + w_{\text{b}}^{-}\:\Gamma _{-+} + w_{\text{b}}^{2}\:\Gamma _{2+} \\
&=& \frac{3+\cos(4\theta)}{2}\: \ga + \frac{5+3\cos(4\theta)}{3+ \cos(4\theta)}\: \gad
\end{eqnarray*}
The width $\Gamma ^{w}_{\text{c}}$ of the narrow peak in first order of $\gad / \ga$ calculates to
\begin{eqnarray}
\Gamma ^{w}_{\text{c}} &=& w_{\text{c}}^{+} \: \Gamma _{+2} + w_ {\text{c}}^{-} \: \Gamma _{-2} + w_{\text{c}}^{2} \: (\Gamma _{2+} + \Gamma _{2-}) \nonumber \\
&=& 2(a + 1)\gad - \frac{5+3\cos(4\theta)}{3+ \cos(4\theta)}\: \gad \label{weight-width}
\end{eqnarray}
These results agree with the exact secular approximation results $\Gamma_{\text{b}}$ and $\Gamma_{\text{c}}$  in first order in $\gad / \ga$, \eqnsref (\ref{sec-exakt-mollow}) and (\ref{sec-exakt-peak}). This justifies our choice for $\alpha _{\text{b}}$ and $\alpha _{\text{c}}$. 

In the secular limit, the result for the width of the additional peak is also in agreement with the corresponding expressions obtained in the analytical section, \eqnsref  (\ref{phys-width}) and (\ref{peak-breite-dauer}). Inserting the explicit values \eqnsref (\ref{sinval}) and (\ref{cosval}) for the trigonometric functions into the rhs. of \eqnref (\ref{weight-width}) yields
\[
\Gamma^{w}_{\text{c}} = \frac{\Omega ^2+2a(2\Delta^2 + \Omega^2)}{2\Delta^2 + \Omega^2} \: \gad
\]
Taking the secular limit in the expression for the peak width \eqnsref (\ref{phys-width}) and (\ref{peak-breite-dauer}) calculated in the analytical section by expanding to zeroth order in $\ga / \Omega$ yields the same result. 
Thus the spring model is an adequate way of visualizing the atomic dynamics responsible for the central spectral features in the secular approximation.

Of course also the satellite peaks in the Mollow spectrum can be discussed in this way. But as there is always only one spectral component at each side, this does not give new insights. 

A corresponding spring model yields the correct results also for a two-level system driven by one laser field. Thus the model is not limited to the three-level system discussed here and could  be useful in predicting the possible modes in more complex atomic systems. 

One might also think of extending the mechanical spring system in a way which allows to drive the oscillations. As discussed in almost any textbook on mechanics, a suitably chosen driving frequency may give rise to resonance effects. Under ideal conditions, it is even possible to excite only one of the modes of the spring model. As each mode corresponds to one of the spectral components in the atomic system, one might hope to eliminate all spectral components but one from the spectrum under these ideal conditions.
There are two different ways to find a suitable spring model. On the one hand, one can try to modify the atomic system, find the corresponding spring model, and then check, whether the modifications allow to drive the spring system. On the other hand, one can alter the spring system such that it allows for an external driving and then try to translate it back to an atomic system. 
However a first attempt to modify the atomic system by an additional laser field on one of the transitions did not succeed to allow for a driving of the spring system. There are also some general arguments against the possibility to excite the various spectral components using this method.  The spring model does not allow one to easily obtain the amplitudes of the respective modes. Therefore it is not clear whether a large excitation of a spring mode corresponds to a large intensity of the corresponding spectral component. Also, the masses and spring constants of the spring model depend on the parameters of the atomic system, and therefore also the resonance frequencies. Thus the parameters of the spring system corresponding to an atomic system continuously change during the excitation process, which is different to conventional spring systems. 
Nevertheless finding an atomic system which translates to a driven spring system  may provide novel insights in the atomic dynamics.

\section{Discussion and conclusions}
In this paper we have investigated the resonance fluorescence spectrum of a three - level atomic system driven by a single monochromatic laser field. The third atomic level adds an additional decay path from the upper laser driven state to the lower one by radiative decays. An additional narrow peak compared to the usual two-level fluorescence spectrum was found. By a numerical analysis, the peak was found to be especially pronounced for a very small coupling of the additional atomic level to the laser driven ones. Also the Rabi frequency of the driving laser must not be too high if the detuning of the driving laser is kept fixed. 

These results were used to derive analytical expressions for the width, the height and the intensity of the additional peak using two different methods. The first method uses a secular approximation in which a suitable choice of the detuning allows to increase the Rabi frequency without losing intensity in the additional peak. In this limit, the peak intensity essentially only depends on the relation $a=\gaz / \gad$ between the decay width from and to the additional level, but not on other variables such as the Rabi frequency $\Omega$ or the absolute value of the decay widths. In addition to the dependence on $a$, the peak amplitude is explicitly proportional to the decay width $\gad$, and the peak width is proportional to $\gad ^{-1}$. In the second method, results from the following discussion of the physical origins of the peak were used to derive an exact expression for the intensity of the narrow peak under the assumption that the random telegraph model is appropriate to describe it. From this expression for the intensity, the width and the amplitude of the peak were calculated under the assumption of a weak coupling of the additional atomic level to the driven ones. In the respective limits, the results of the two different methods are equal.

In the second part of the paper, we have discussed the physical origin of the additional narrow peak. We have applied the random telegraph model to our atomic system. In previous works, this model was used to explain the exhibition of similar additional spectral features in a different atomic system. In this alternative system, the three atomic levels are coupled by two lasers in a V - configuration instead of the additional decay path in our system. The average length of light and dark periods, which are the crucial variables in this model, were calculated analytically. Using these results, we have explained the relation between the atomic system discussed here, the atomic system where both transitions are driven by laser fields, and an experiment which was recently reported. In the experiment, a setup as described in this paper
was used. The experimental results are in quantitative agreement with the theoretical predictions from the random telegraph model. 

Finally we have discussed a simple spring model to explain the atomic dynamics in the secular limit. This model provides an intuitive picture for the different spectral contributions in the fluorescence spectrum and allows to visualize the atomic dynamics responsible for the fluorescence spectra. Also it can be used to easily calculate approximate expressions for the widths of the different peaks.

\begin{acknowledgments}
Funding by Deutsche Forschungsgemeinschaft (Nachwuchsgruppe within SFB 276) is gratefully acknowledged. We would like to thank Prof. Lorenzo M. Narducci for enlightening calculations and discussions in the initial phase of this project and many useful hints on the \nopagebreak final version of the manuscript.
\pagebreak
\end{acknowledgments}


\end{document}